\documentclass[12pt]{iopart}
\usepackage{graphicx}
\usepackage{xspace}
\usepackage{subfigure}

\newcommand{\sqrts}{$\sqrt{s_{_{NN}}}$}

\newcommand{\Npart}{$\langle N_{part}\rangle$\xspace}

\newcommand{\pp}{{\it p-p}\xspace}

\begin{document}

\title[Strange hadrons as dense matter probes]{Using strange hadrons as probes of dense
matter}

\author{Helen Caines}

\address{Physics Department, Yale University, New Haven, CT 06520, U.S.A}
\ead{helen.caines@yale.edu}
\begin{abstract}

 The spectra of strange hadrons have been measured in detail as a function of
centrality for a variety of collision systems and energies at RHIC. Recent results
 are presented and compared to those measured at the SPS. The effects of the
system size  on strange particle production and kinematics are examined.  I  place
specific emphasis on comparing A-A to \pp production and discuss how
strangeness can be used to probe the dense matter produced in
heavy-ion collisions.

\end{abstract}


\section{Introduction}

A wealth of data has and is being collected at the RHIC accelerator, BNL. The STAR experiment~\cite{STAR}, by dint
of its large acceptance for charged hadrons at mid-rapidity,
is ideally suited for measuring strange hadron production. I therefore concentrate in my talk on results from STAR
and contrast these to those from the SPS, CERN.

The unprecedented  detail of the RHIC measurements, due to the high
statistics of the data taken, allows us to probe the centrality dependence of the produced particles. The
fact that RHIC can  be operated over a range of collision energies and species means that the same detectors can be utilised
to perform systematic scans. Since RHIC is a collider, the  experiments
 obtain their measurements over the same phase space no matter what the collision species or energy (except for the rare
case of asymmetric collisions). This, so far, unique feature of RHIC  removes
issues arising from differing experimental trigger and reconstruction biases when  contrasting results from different
experiments.

\section{Statistical models}\label{Section:StatModel}

In ultra-relativistic heavy-ion collisions, the important question of chemical equilibration of the produced matter
remains open. While there is much evidence suggesting that the matter is equilibrated, there is, as yet, no conclusive proof.
The detailed study of strange hadron production, including multi-strange baryons, is key to unlocking the answer to this
and other questions. One standard method for detecting chemical equilibration is via statistical models. These
models aim to extract the degree of chemical equilibration of the matter, along with the freeze-out temperature, T$_{ch}$.
 This is done via contrasting the experimentally measured particle ratios to those calculated from the models assuming a source
 in thermal equilibrium. Such models cannot prove equilibration, since the agreement may be coincidental, however if the
 measured ratios cannot be reproduced, the source cannot be in equilibrium at the point when the hadronic ratios are fixed.

 Many such models have been applied to
heavy-ion collisions, however, there are three that are readily
available to experimentalists. These are, THERMUS~\cite{Thermus},
SHARE~\cite{Share} and a four parameter fit model~\cite{Kaneta}. All three
models have T$_{ch}$, $\mu_{q}$, $\mu_{s}$ and $\gamma_{s}$ as free
parameters. The latter term accounting for  incomplete strangeness chemical
equilibration. SHARE and THERMUS include
a chemical potential to account for charge/isospin conservation, a
charm chemical potential and $\gamma_{c}$. SHARE alone allows for a
$\gamma_{q}$ while THERMUS can perform both Grand Canonical and
Canonical ensemble calculations.

Before attempting to use these models to fit the STAR Au-Au \sqrts =
200 GeV data using all their free parameters, a consistency check was
made. This ensured that the same results were obtained when the models
were constrained by a Grand Canonical calculation using only
T$_{ch}$, $\mu_{B}$, $\mu_{s}$ and $\gamma_{s}$. While the results were
consistent within errors, a more detailed look at the models is
underway in order to understand why the results are not in complete agreement.

\subsection{Fits to Central Au-Au and minimum bias \pp results.}

\begin{figure}[htb]
\centering
\mbox{\subfigure[Grand Canonical]{\includegraphics[width=0.45\textwidth]{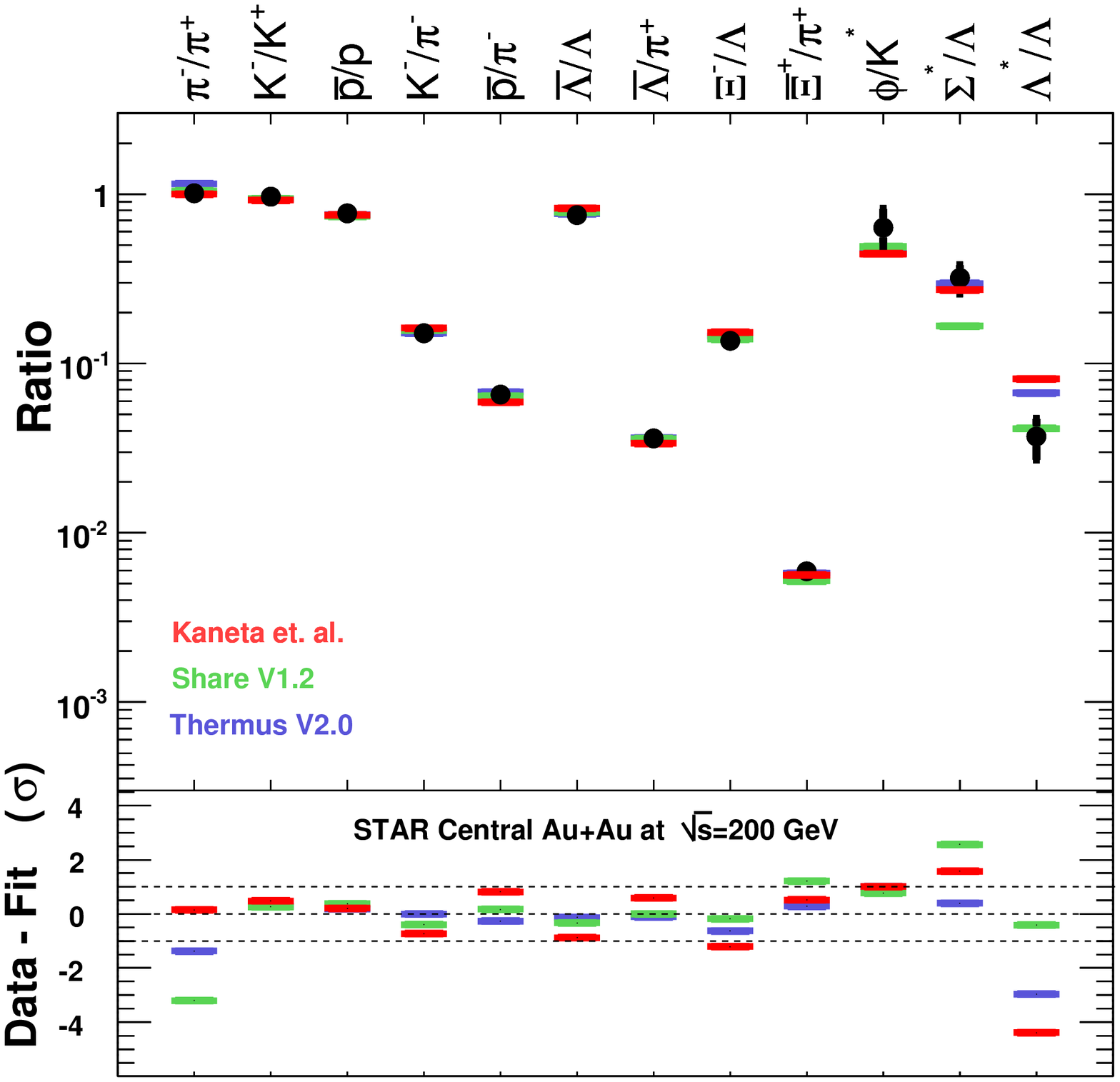}}\quad
\subfigure[Canonical]{\includegraphics[width=0.45\textwidth]{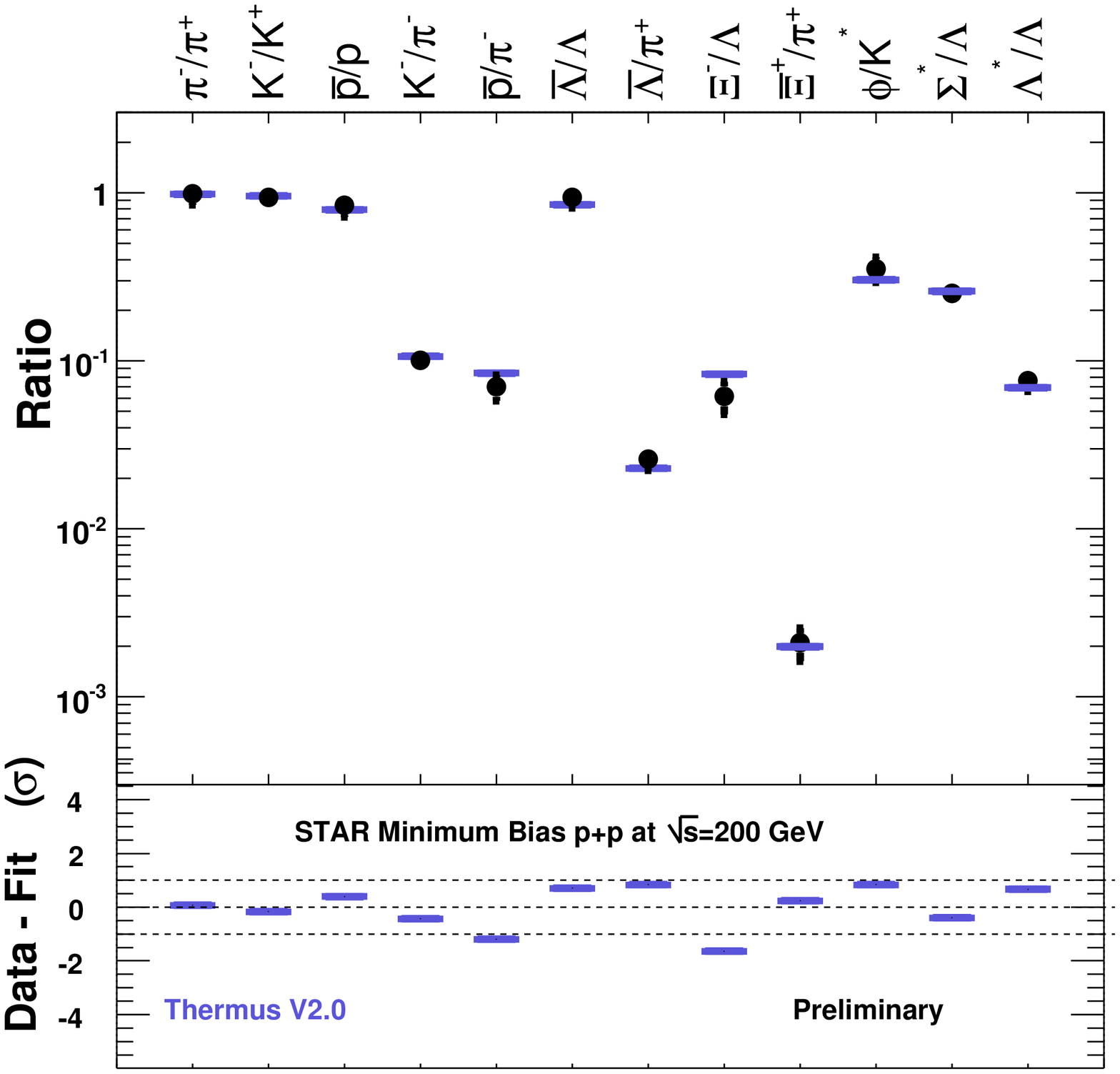}}}
\caption{Statistical model fits to preliminary STAR data a) 0-5$\%$ most central Au-Au
and b) \pp, \sqrts = 200 GeV. Solid points represent STAR preliminary
data the coloured bars the calculations. The bottom panel represents
the difference between the data and the calculations for each
ratio.} \label{Fig:Chempp} \label{Fig:ChemAuAu}
\end{figure}

Figure \ref{Fig:ChemAuAu}a and Table~\ref{Table:AuAuFit} present the results of the fits to the
0-5$\%$ most central preliminary Au-Au data from STAR~\cite{ChemFit}. It can be
seen (bottom panel of the figure) that the stable particle ratios, with the exception of the pions,
are well represented by each of the models. The data for the
resonances, on the other hand, are less well produced.

 While each model describes the data, there are
significant deviations between the results. All models
calculate that the baryon and strangeness chemical potentials are close
to zero yet there is a 30 MeV difference in the chemical freeze-out
temperature, this difference being several sigma beyond the models' calculated
errors. THERMUS and the 4 parameter fit show strangeness saturation,
$\gamma_{s}$ =1, while SHARE calculates an over-saturation with
$\gamma_{s}$ =2.1 $\pm$ 0.6. Within the SHARE framework, the light
quarks are also over saturated with $\gamma_{q}$ =1.7 $\pm$ 0.5. The
variations in these results should be taken as the degree of
systematic uncertainty in the chemical freeze-out parameters determined
 for  RHIC heavy-ion collisions. It should be noted however,
that all the models determine that, when the stable particle ratios are fixed,
the medium produced at RHIC is close to the
critical temperature for transition to a QGP, as predicted by lattice
QCD~\cite{Lattice}.

\begin{table}[hb]
\begin{center}
\begin{tabular}{|c|c|c|c|c|c|}
\hline

   Model & T$_{ch}$ (MeV) & $\mu_{B}$ (MeV)& $\mu_{S}$ (MeV)& $\gamma_{s}$  & $\gamma_{q}$\\
  \hline
  THERMUS & 168 $\pm$ 6  &   45 $\pm$ 10   &  22 $\pm$ 7        & 0.92 $\pm$ 0.06 & N/A\\
  \hline
  SHARE   & 133 $\pm$ 10    & 23  $\pm$ 19   & 5  $\pm$ 7  & 2.0 $\pm$ 0.6 & 1.7 $\pm$ 0.5\\
  \hline
  4 Parameter & 161 $\pm$ 5     &  23 $\pm$ 7  & 8  $\pm$ 5     & 1.01 $\pm$ 0.06 & N/A\\

 \hline
 \end{tabular}

\caption{Results from the models described in the text of Grand
Canonical ensemble fits to Au-Au \sqrts = 200 GeV preliminary data from STAR.}
\label{Table:AuAuFit}
\end{center}
\end{table}
 The above results should be contrasted with those of a fit
to preliminary STAR \pp data at \sqrts = 200 GeV as shown in
Fig.~\ref{Fig:Chempp}b. For this fit, THERMUS is used.  It is
currently the only readily available model that can perform a
canonical ensemble fit. Again, the good quality of the fit should be noted. It is
also interesting to observe that in the \pp data the resonance ratios are  well
represented. For this fit, T$_{ch}$ = 171 MeV $\pm$ 9 MeV,
close to that arrived at for the Au-Au case. However, $\gamma_{s}$ =
0.53 $\pm$ 0.04, indicating that the \pp data are far from strangeness
saturation.

\section{Centrality dependence of strangeness yields}

Since there is a marked difference in the composition of the
hadrons produced in \pp and central Au-Au collisions it is
necessary to study the centrality dependence of strangeness
production. In this way it can be determined if the transition to strangeness
saturation occurs abruptly at a given
centrality or changes smoothly.

\subsection{Strangeness enhancement}

One such study is performed using strange (anti)baryons
and the enhancement factor, E(i), calculated
as
\begin{equation}
 E(i) = \frac{Yield^{AA}(i)*\langle N_{part}^{NN}\rangle}{Yield^{NN}(i)*\langle N_{part}^{AA}\rangle}.
 \label{Eqn:Enhance}
\end{equation}

\noindent If the yields of the hyperons scale with \Npart, E(i) will be
unity over all centralities. Such a scaling is predicted in the Grand Canonical regime when
strangeness is fully equilibrated.

\begin{figure}[htb]
\centering
\mbox{\subfigure[]{\includegraphics[width=0.45\textwidth]{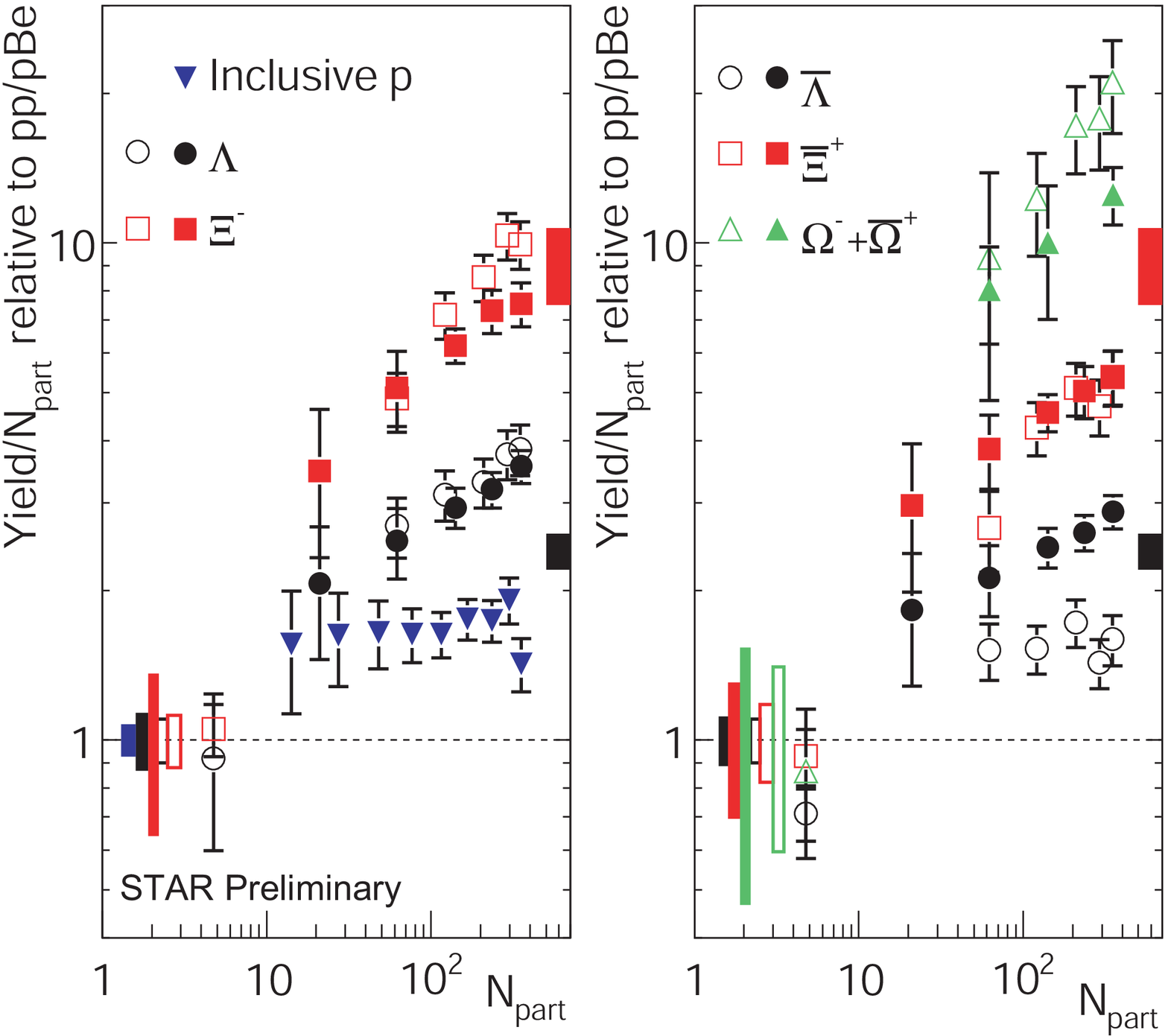}}\quad
\subfigure[]{\includegraphics[width=0.45\textwidth]{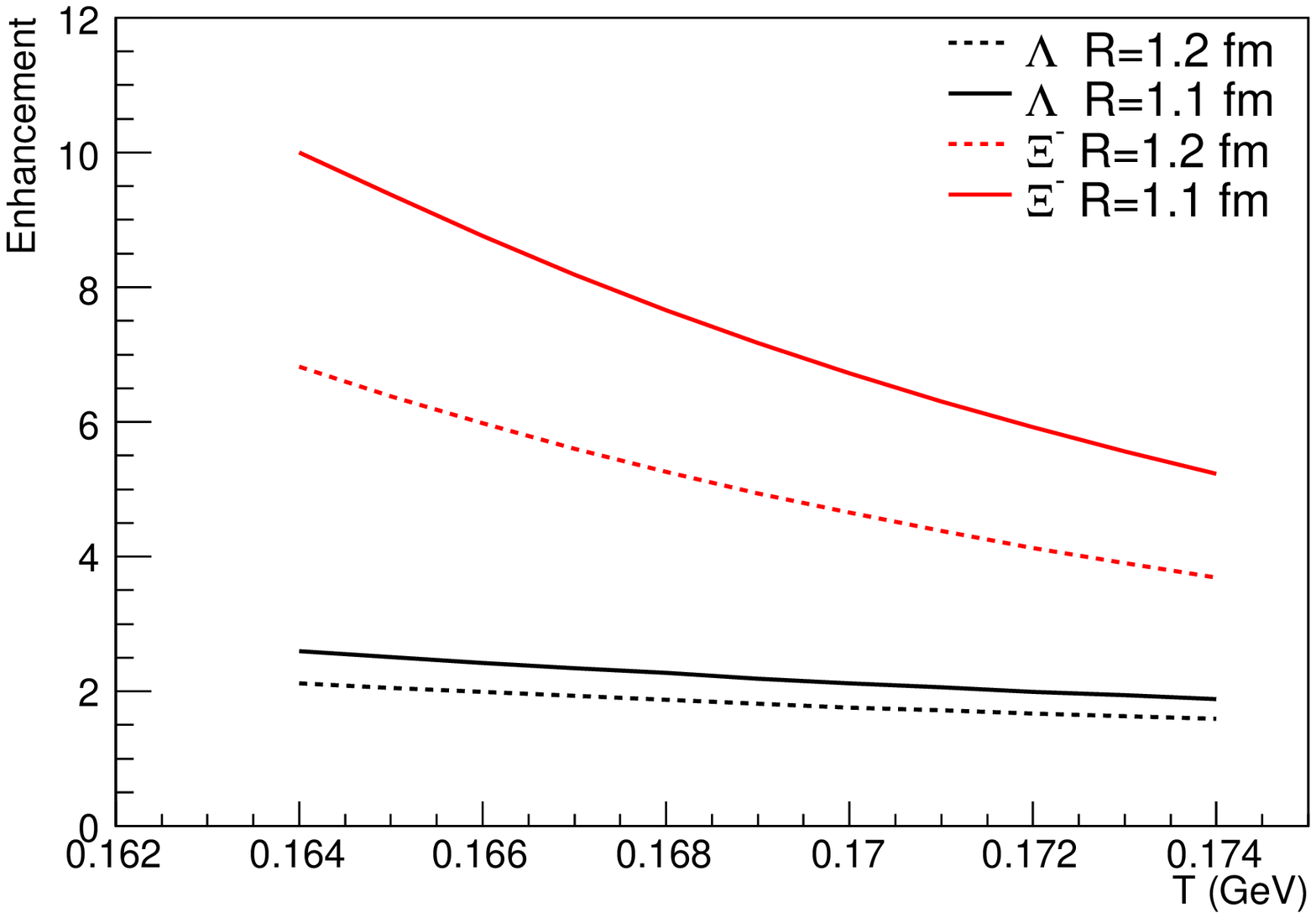}}}
\caption{ a) Enhancement factors as a function of \Npart for strange
(anti)baryons. Solid symbols are for Au-Au collisions at \sqrts = 200 GeV,  and the
open symbols are measured by NA57 from Pb-Pb at \sqrts=17.3 GeV. The
boxes represent the combined statistical and systematical
uncertainties in the \pp and p-Be data. The error bars on the
data points represent those from the heavy-ion measurement. The boxes on the
right axes mark the predictions from a model using a Grand Canonical
formalism. b) Calculated enhancements using the
model described in \cite{Redlich} as a function of chemical
freeze-out temperature. Two values of R$_{0}$ are assumed to calculate the
strangeness production correlation volume.} \label{Fig:EnhanceComp} \label{Fig:EnhanceT}
\end{figure}

Figure \ref{Fig:EnhanceComp}a shows the results of this calculation at
RHIC~\cite{Matt} compared to those from NA57 at the SPS for Pb-Pb collisions at
\sqrts = 17.3 GeV~\cite{NA57Enhancea,NA57Enhanceb}. Also shown are
the results for STAR's measured inclusive protons, which include all
feed-down protons from hyperon decays. It can be seen that the enhancement
increases with increasing strangeness content and that, at RHIC, the
anti-baryon enhancements are approximately those of the baryons.
This reflects the near zero net-baryon density at RHIC. While the
enhancements are similar between RHIC and SPS data, the magnitude of
the normalization uncertainty, reflected by the error boxes at
\Npart= 1 should be noted. This normalization uncertainty is due to the error in the \pp and p-Be
data and can only be reduced by improving the event statistics for
these collisions.

 The solid boxes on the right hand axis of Fig.~\ref{Fig:EnhanceComp}a
represent the predicted  enhancements using a statistical
model calculation. This model uses a canonical regime calculation for \pp
 and assumes that the Grand Canonical regime is reached in central
A-A collisions~\cite{Redlich}. The bottom edge of the boxes  uses a
chemical freeze-out temperature, T$_{ch}$, of 170 MeV for
both \pp and Au-Au collisions. The top of the boxes represents
the same model calculation for T$_{ch}$ = 165 MeV. This range of temperatures
is within the uncertainty of the chemical freeze-out temperature obtained in these collisions, as discussed
in section~\ref{Section:StatModel}. The predicted enhancement
decreases with increasing freeze-out temperature since the
enhancement within this model is due to  reduced phase space for
strangeness production in \pp. The available phase space
increases with temperature and volume of the produced fireball and hence the
enhancement decreases if either term increases. It appears from this calculation that
  the model cannot differentiate between the two temperatures.

 However, as stated, the enhancement is sensitive to both the temperature of the system and
its volume. This volume term is proportional to the cube
of the proton radius, R$_{0}$. This cubic dependency implies that the enhancement factors are
very sensitive to the assumed radius,
Fig.~\ref{Fig:EnhanceT}b ~\cite{RedlichPrivate}. This dual dependency
on T$_{ch}$ and R$_{0}$ means that a given enhancement can be
obtained  via two combinations, the multi-strange baryons
having the greatest sensitivity.

\begin{figure}[htb]
\centering
\mbox{\subfigure[]{\includegraphics[width=0.45\textwidth]{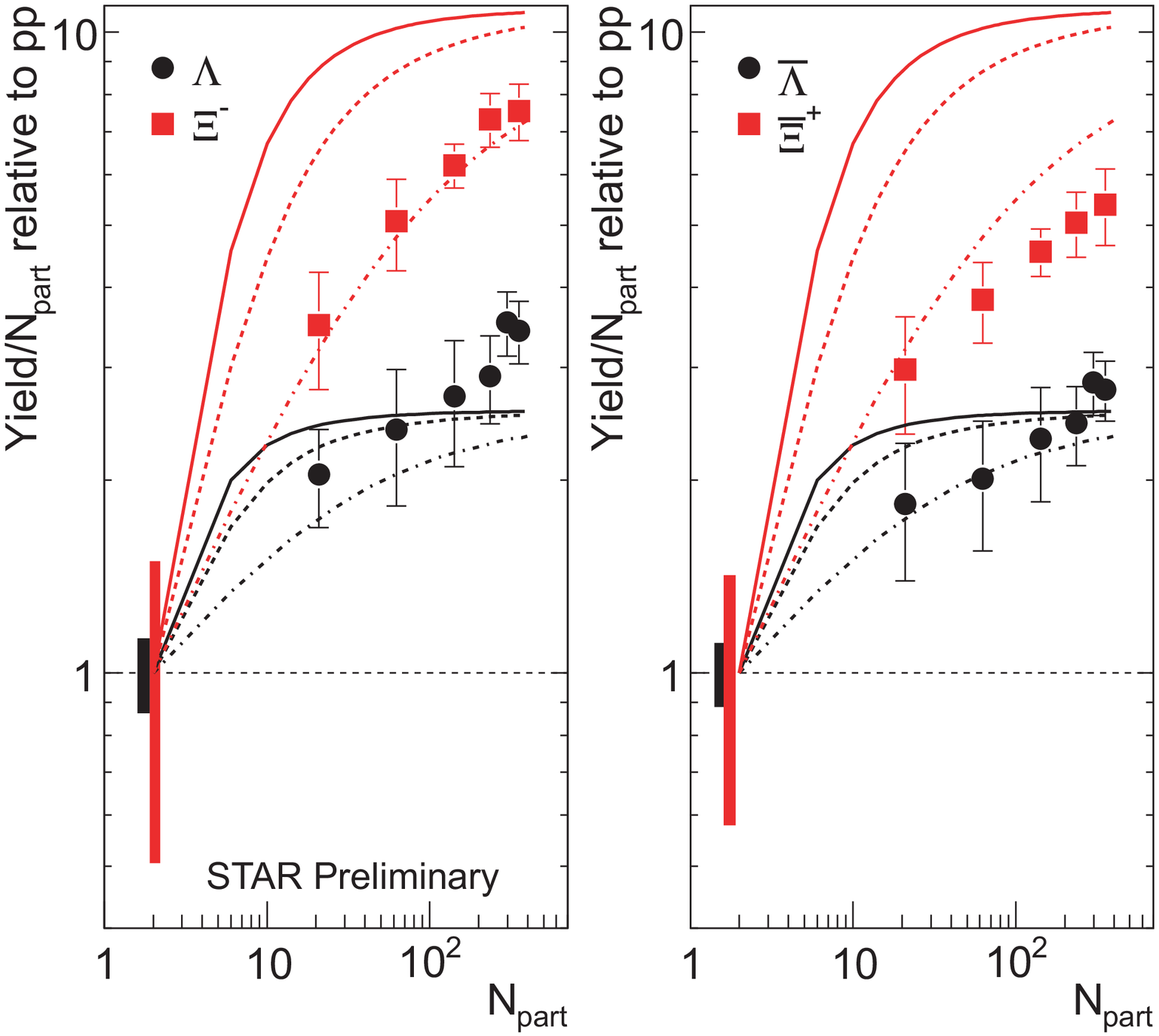}}\quad
\subfigure[]{\includegraphics[width=0.45\textwidth]{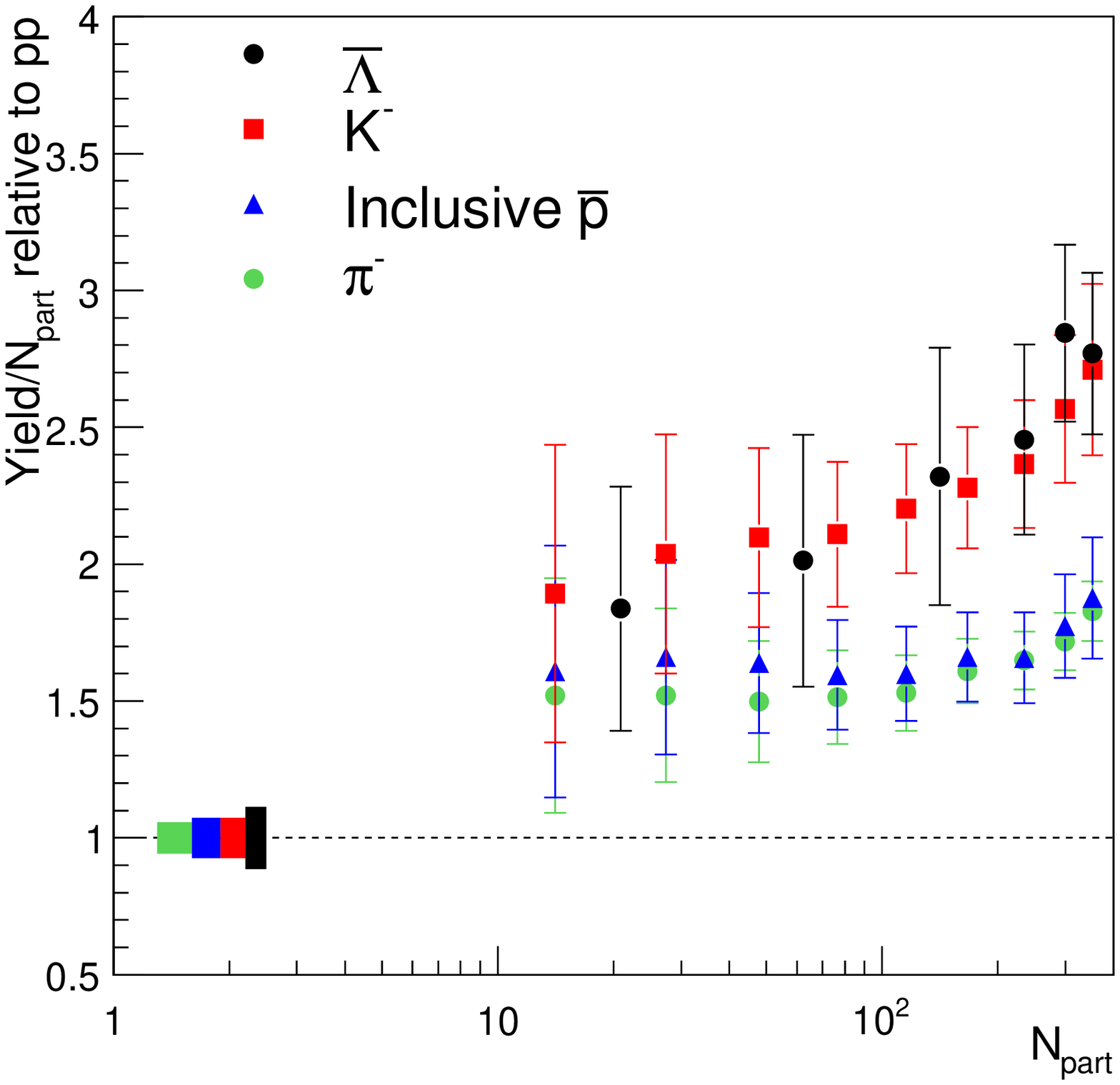}}}
\caption{a) Enhancement factors as a function of \Npart for strange
(anti)baryons.  Errors as explained in Fig.~\ref{Eqn:Enhance}a). The curves are for different assumptions of the \Npart
dependence of the strangeness suppression. The solid curve is \Npart
scaling, dashed curve - $\langle N_{part}\rangle^{2/3}$, dot-dashed curve -
$\langle N_{part}\rangle^{1/3}$. A common T$_{ch}$ of 165 MeV is assumed.
b) The enhancements factor calculated for preliminary data from STAR for identified particles.
The data are from STAR for Au-Au collisions at \sqrts = 200 GeV. } \label{Fig:EnhanceCurve}\label{Fig:Ncharge}
\end{figure}

 It is believed that there is a linear correlation
between the geometric overlap volume of the collision and \Npart. The strangeness
production correlation volume and \Npart were also expected to be linearly related.
However, when a comparison of the { \it shape} of
the enhancement as a function of \Npart to that predicted by the model assuming a linear
term in \Npart is made (see the solid curve in Fig.~\ref{Fig:EnhanceCurve}a) the
measured  functional form is not well represented. Figure \ref{Fig:EnhanceCurve}a also shows
a $\langle N_{part}\rangle^{2/3}$, dashed curve, and $\langle N_{part}\rangle^{1/3}$, dashed-dot curve. The best
description of the data, in both shape and magnitude, is achieved when a
$\langle N_{part}\rangle^{1/3}$ scaling is used in combination with T$_{ch}$ = 165 MeV and a radius R$_{0}$ = 1.1 fm.

\subsection{Flavour dependence of the scaled yields}

When studying the enhancement for strange baryons it is helpful to also examine the possible enhancement
of non-strange baryons and mesons. Figure \ref{Fig:Ncharge}b shows the results for identified $\pi^{-}$, K$^{-}$
and inclusive, i.e. including all feed-down, anti-protons~\cite{PIDyield} compared to $\bar{\Lambda}$ at \sqrts = 200 GeV. It can
be seen that there is an enhancement for all species and that the enhancement scales with the number of (anti-)strange
quarks in the particle. While it is expected that the strange quarks``see" a phase space suppression in peripheral
A-A and \pp collisions the light quarks are not expected to be affected in such a manner. Since the models do
not predict an enhancement for anti-protons and pions, any phase-space suppression effects for strangeness
should be made relative to the observed increase in the measured pion yields.

\section{Beyond the bulk}

So far I have dealt only with integrated yields, a measure dominated by the low p$_{T}$ region, p$_{T} < $ 1 GeV/c.
To probe the particle production further, we study the nuclear modification factor, R$_{AA}$. The nuclear modification
 factor compares p$_{T}$ spectra from A-A collisions with binary scaled \pp data. When \pp results are not available
 an alternative measure, R$_{CP}$, is calculated by substituting binary scaled peripheral A-A results for the \pp spectra.
When using R$_{CP}$, it is assumed that the peripheral A-A data are a close approximation to \pp at the same energy. However,
in the previous sections I have shown that the \Npart scaled central yields of $\Lambda$ and $\Xi$ are enhanced by factors of
 3 and 7 respectively, compared to \pp. Even the most peripheral \Npart scaled data are a factor of 2-3 enhanced. Thus,
 the assumption that
peripheral data can be substituted for \pp is clearly incorrect. However, a study of both measures remains informative.

\begin{figure}[htb]
\centering
\mbox{\subfigure[R$_{CP}$ for identified particles.]{\includegraphics[width=0.45\textwidth]{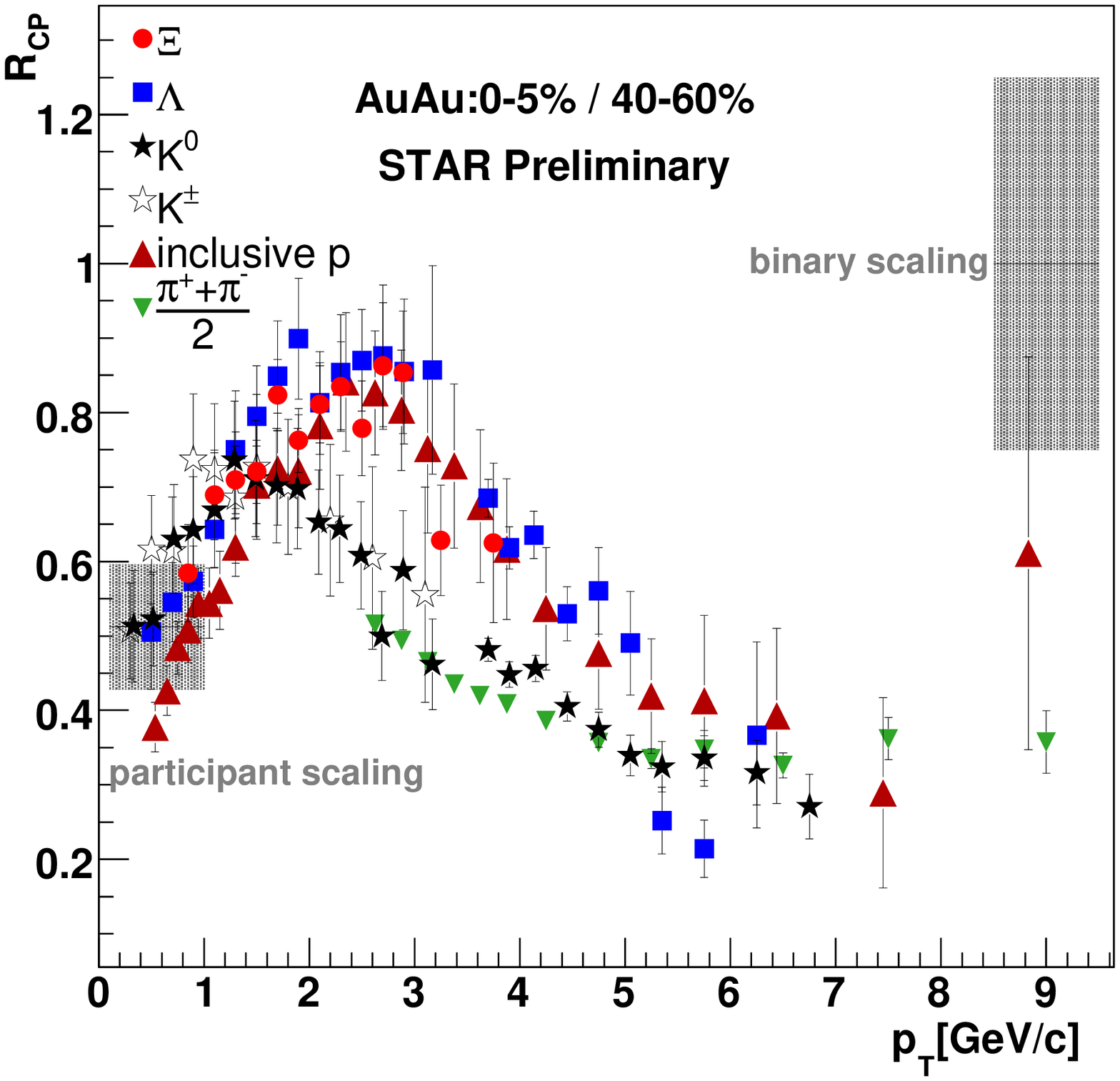}}\quad
\subfigure[R$_{CP}(\Lambda)/R_{CP}(K^{0}_{S}$).]{\includegraphics[width=0.5\textwidth]{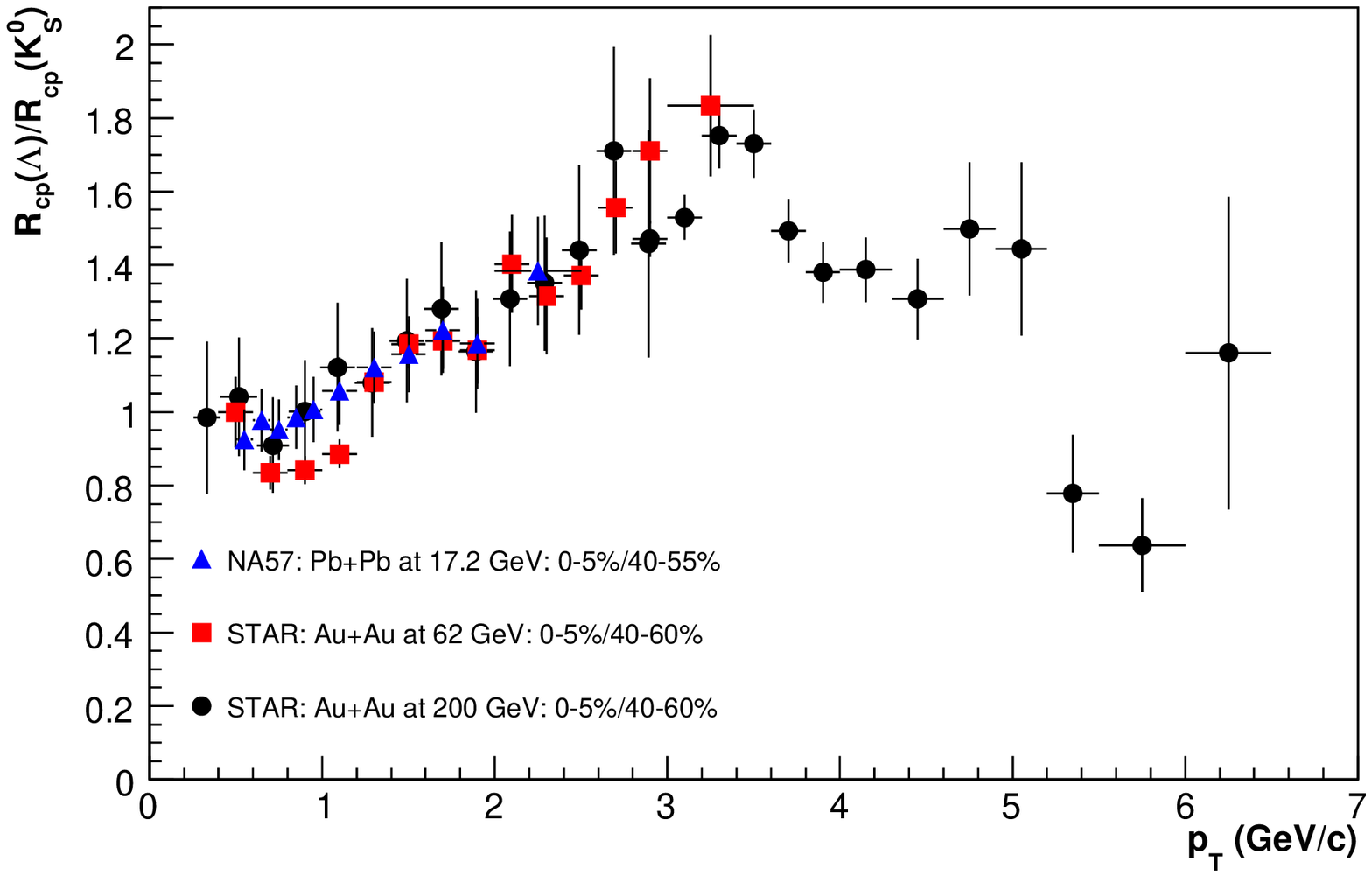}}}
\caption{a) R$_{CP}$ for Au-Au collisions at \sqrts = 200 GeV. b) The ratio of the R$_{CP}$($\Lambda$) over
 R$_{CP}(K^{0}_{s}$) for different collisions energies from RHIC and the SPS.} \label{Fig:Rcp}\label{Fig:RcpRatios}
\end{figure}

 The measured R$_{AA}$ for 0-5$\%$ Au-Au for $\Lambda$, $\Xi$ and inclusive protons has been shown previously~\cite{Raa}.
  This data show that  R$_{AA}(\Xi) > R_{AA}(\Lambda) > R_{AA}(\bar{\rm p}) > R_{AA}(K^{0}_{S})$ at intermediate p$_{T}$. All
particles except the kaon peak above unity. The R$_{CP}$ measure, on the other hand, shown in Fig.~\ref{Fig:Rcp}a, shows
R$_{AA}(\Xi)$ = R$_{AA}(\Lambda)$ = R$_{AA}(\bar{\rm p})$ and an R$_{CP}$ maximum of approximately 0.8~\cite{Matt}. The inclusion of the
kaon and $\pi$ in Fig.~\ref{Fig:Rcp}a also reveals a baryon/meson splitting up to p$_{T}$ values of $\sim$ 6 GeV/c, after which
 a constant suppression at 0.2 is observable for all species. This suppression with respect to binary scaling is taken as
 evidence of energy loss of partons as they pass through the hot medium produced in central Au-Au collisions at
 RHIC~\cite{JetQuenchTheory1,JetQuenchTheory2}. Meanwhile the baryon/meson splitting at
 intermediate p$_{T}$ can be explained via quark recombination models~\cite{Reco}. The observed  suppression of high p$_{T}$ particles
 in central Au-Au events and the baryon/meson splitting at intermediate
 p$_{T}$ are strong indications that the system enters a partonic phase with a large gluon density. The striking difference between the
 R$_{CP}$ and R$_{AA}$ measurements can be understood in terms of the large phase space suppression effects, which
 dominate the bulk measurements, extending into the high p$_{T}$ domain. The R$_{AA}$ is therefore a convolution of two effects with the
 phase space suppression dominating. However, in  peripheral events, the phase space effects are already significantly diluted
  but those of the hot dense medium are not yet large.   R$_{CP}$, for strange particles, is therefore a more sensitive measure
  of ``jet quenching" effects at intermediate and high p$_{T}$.

  R$_{CP}$ measurements have also been made in Au-Au collisions at \sqrts = 62 GeV at RHIC~\cite{Raa} and at the SPS for
 \sqrts = 17.3 GeV~\cite{SpsRcp}. In both cases, there is a clear  splitting between the $\Lambda$ measurements and the K$^{0}_{S}$
 measurements in the intermediate p$_{T}$ regime. While the peaks of all the measurements occurs at approximately the same p$_{T}$
 for all three measurements  the value of the  maxima  increases as the collision energy decreases.
 The peak of the SPS measurements have a value of $\sim$ 1.3 and 1 for $\Lambda$ and K$^{0}_{S}$ respectively while the
 top RHIC values are $\sim$ 0.9 and 0.7. The splitting of the baryons
 and mesons however appears to be similar. To probe this probability further, we plot in Fig.~\ref{Fig:RcpRatios}b the R$_{CP}(\Lambda)$/
 R$_{CP}(K^{0}_{s})$ ratio. It is striking that the shape of this ratio is the same for each of the three energies. This
  suggests that any conclusions drawn from the splitting of baryons/mesons at \sqrts= 200 GeV should also be concluded for
  top SPS energies. It may however be accidental but that would require three different sets of mechanisms happening to sum
  to the same result which is unlikely.

\section{Summary}

In summary, in central Au-Au collisions at energies of \sqrts = 200 GeV,  we observe a distinct enhancement of strangeness production compared to
that in \pp collisions at the same energy. If the origin of the measured enhancement is due to a lack of available
phase space in \pp and peripheral collisions, then these phase space effects diminish with an apparent
$\langle N_{part}\rangle^{1/3}$ dependency. The effects of this enhancement can be observed into
the intermediate p$_{T}$ regime.

 A clear suppression of intermediate and high p$_{T}$ particles is observed. It can be explained via models assuming partonic
energy loss. Phase space suppression effects make the R$_{CP}$ ratio a cleaner measurement from which to extract
``jet quenching" effects since R$_{AA}$ is a convolution of phase space suppression and partonic energy loss effects.
The baryon/meson splitting at intermediate p$_{T}$ is seen at all energies from \sqrts = 17.3 to 200 GeV. Although
the scale of the suppression is reduced at lower energies the slope of fractional difference between the $\Lambda$ and K$^{0}_{S}$
R$_{CP}$ values is the same. This suggests that recombination is present at both RHIC and top SPS energies.

\subsection*{Acknowledgments}

I wish to thank Krzysztof Redlich for very helpful discussions and
providing many of the theoretical curves. Thanks also go to Sevil Salur, Georgio Torrieri, Ingrid Kraus, and Anton Andronic
for working with me to investigate the various statistical models and their differences.


\section*{References}

\begin{thebibliography}{10}

\bibitem{STAR} M.~Anderson {\it et al.} (STAR Collaboration), Nucl. Instrum. Meth. A {\bf 499}, 659 (2003).

\bibitem{Thermus} S.~Wheaton and J.~Cleymans, hep-ph/0407174 (2004).

\bibitem{Share} G.~Torrieri {\it et al.}, Comput. Phys. Commun {\bf 167} 229 (2005).

\bibitem{Kaneta} N.~Xu and M.~Kaneta, Nucl. Phys. {\bf A698} 306c (2002).

\bibitem{ChemFit} S.~Salur, (for the STAR Collaboration)  nucl-ex/0606006 (2006).

\bibitem{Lattice} F.~Karsch J. Phys. G {\bf 31} 641 (2005).

\bibitem{Matt} M.A.C.~Lamont  (for the STAR Collaboration), these proceedings.

\bibitem{NA57Enhancea} E.~Andersen {\it et al.} (WA97 Collaboration),
Phys. Lett. {\bf 449} 401 (1999).

\bibitem{NA57Enhanceb} F.~Antinori  (for the WA97/NA57 Collaboration),
Nucl. Physics {\bf A698} 118 (2002).

\bibitem{Redlich} A.~Tounsi, A.~Mischke and K.~Redlich,
Nucl. Phys. {\bf A715} 565 (2003).

\bibitem{RedlichPrivate} K.~Redlich private communication, based on
method described in \cite{Redlich}.

\bibitem{PIDyield} J.~Adams {\it et al.} (STAR Collaboration), Phys. Rev. Lett. {\bf 92} 112301 (2004).

\bibitem{Raa} S.~Salur, (for the STAR Collaboration) QM2005 nucl-ex/0509036 (2005).

\bibitem{JetQuenchTheory1} M.~Gyulassy and M.~Plumer, Phys. Lett. B {\bf
243} 432 (1990); R.~Baier {\it et al. ibid} {\bf 345}, 277 (1995).

\bibitem{JetQuenchTheory2} X.N.~Wang and M.~Gyulassy, Phys. Rev. Lett. {\bf
68} 1480 (1992); X.N.~Wang Phys. Rev. C {\bf 58}, 2321 (1998).

\bibitem{Reco} R.J.~Fries {\it et al.} Phys. Rev. C {\bf 68} 044902 (2003),
V.~Greco {\it et al.} Phys. Rev. C {\bf 68} 034904 (2003) and
R.C.~Hwa {\it et al.} Phys. Rev. C {\bf 67} 034902 (2003).

\bibitem{SpsRcp} F.~Antinori {\it et al.} (NA57 Collaboration) Phys. Lett. {\bf B 623} 17 (2005).

\end{thebibliography}
\end{document}